\documentclass[showpacs, pra,twocolumn,preprintnumbers ,amsmath, amssymb, superscriptaddress, aps]{revtex4-2}
\usepackage{color}
\usepackage{amsmath,amssymb}
\usepackage{pifont}
\usepackage{amssymb}  
\usepackage{bbold}
\usepackage{float}
\usepackage{subfloat}
\usepackage[caption=false]{subfig}
\usepackage{tikz}

\usepackage{makecell}
\usepackage{subfig}
\usepackage{pifont}   
\usepackage{graphicx} 
\graphicspath{{Figures/}}
\usepackage{dcolumn}  
\usepackage{bm}       
\usepackage{multirow} 
\usepackage{placeins}
\usepackage[colorlinks=true,linkcolor= red,citecolor=green,]{hyperref}
\usepackage{mathtools}
\usepackage[title]{appendix}

\begin{document}
	\title{Quantum transport in gapped graphene under strain and laser–electrostatic barriers}
\date{\today}
\author{Hasna Chnafa}
\email{chnafa.hasna@gmail.com}
\affiliation{Laboratory of Theoretical Physics, Faculty of Sciences, Choua\"ib Doukkali University, PO Box 20, 24000 El Jadida, Morocco}

\author{Clarence Cortes}
\affiliation{Vicerrector\'ia de Investigaci\'on y Postgrado, Universidad de La Serena, La Serena 1700000, Chile}  
\author{David Laroze}
\affiliation{Instituto de Alta Investigaci\'on, Universidad de Tarapac\'a, Casilla 7D, Arica, Chile}
\author{Ahmed Jellal}
\email{a.jellal@ucd.ac.ma}
\affiliation{Laboratory of Theoretical Physics, Faculty of Sciences, Choua\"ib Doukkali University, PO Box 20, 24000 El Jadida, Morocco}


\begin{abstract} 
	Electron transport in graphene under a laser-modulated barrier is studied in the presence of an energy gap, a scalar potential, and a uniaxial zigzag strain. The transfer-matrix approach is used with the boundary conditions to derive the transmission probabilities as functions of different system parameters. Without strain, raising either the energy gap or the potential generally reduces transmission in the central and lower sidebands. Moderate zigzag strain generates pronounced Fano-type oscillations that vanish at large strain,  while transmission increases for low potential and decreases for high values. In the upper sideband, the incidence energy shifts the resonance peaks to the right, and growing the barrier width generates characteristic oscillatory patterns. Furthermore,  increasing the laser field amplitude enhances transmission, whereas higher laser frequencies tend to suppress it. These findings offer new perspectives on controlling electronic transport in gapped graphene via external fields, strain, and potential applications in optoelectronic devices.
\end{abstract}	
\maketitle


\section{Introduction}
{\color{black}Since its discovery \cite{Nov1,Novoselov}, graphene has established as the ultimate two-dimensional monolayer material, recognised for its exceptional electronic and mechanical properties \cite{castro}. These include phenomena such as Klein tunneling \cite{Nov3}, the anomalous quantum Hall effect \cite{Novoselov,Zhang}, specular Andreev reflection \cite{Beenakker}, unusually high charge carrier mobility \cite{Bolotin}, remarkable thermal conductivity \cite{Balandin}, and mechanical strength \cite{Guinea,Levy,Pereira,Guinea1,Sturla,Oliva}, and many others.  However, as pristine graphene has no band gap, its direct use in electronic and optoelectronic devices such as field-effect transistors and solar cells is limited. To make it suitable for these applications, it is essential to both induce a band gap and precisely control its width. This requirement has motivated extensive research, leading to the development of various practical methods to effectively open and adjust the band structure.  Among these methods are {\color{black}quantum confinement in nanoribbons} \cite{Ritter,Todd}, which creates a band gap by confining the movement of electrons within thin strips of graphene, with the size of the band depending on the width of the ribbon and the geometry of its edges.  {\color{black}Chemical functionalisation} \cite{Shih,Shi,Niyogi}, including oxidation, hydrogenation, or fluorination also opens a band gap by breaking the lattice symmetry, converting certain carbon atoms from $sp^{2}$ to $sp^{3}$ hybridisation.  {\color{black}Another strategy consists in modifying the number of graphene layers \cite{Zhang1,Ohta}: in bilayer or multilayer configurations}, applying a perpendicular electric field breaks inversion symmetry and produces a controllable band gap.  {\color{black}Substrate engineering also plays a crucial role} \cite{Giovannetti,Zhang2}. When graphene is placed on specific substrates, notably hexagonal boron nitride, the slight lattice mismatch and the difference in on-site energies between boron and nitrogen atoms generate a shifted sublattice potential within the graphene layer \cite{San}. This potential breaks the equivalence between carbon sublattices A and B, lifting the symmetry of the sublattice. As a result, the Dirac points are no longer degenerate, giving rise to a finite band gap in the electronic spectrum.}

Apart from these methods, some studies have explored alternative ways to tune the electronic properties of graphene. {\color{black}Mechanical strain} represents another effective approach to alter the electronic structure and potentially create a band gap \cite{Levy,Cocco,Choi}. When strain is applied, the graphene lattice becomes distorted, causing the Dirac cones at the $K$ and $K'$ points to move in opposite directions and modifying the Fermi velocity \cite{Huang}. The effect of strain depends strongly on both its magnitude and direction. Uniaxial or biaxial tension or compression along the armchair or zigzag directions can respectively decrease or increase the Fermi velocity \cite{Wong}. Theoretical studies indicate that strains above $20\%$ might be needed to open a significant band gap \cite{Cocco,Choi,Pereira1}, although experimentally, achievable strains are usually around $1\%$ due to limited transfer from the substrate. Remarkably, strains up to $30\%$ have been observed in suspended graphene, indicating the potential for larger band-gap tuning \cite{Lee}. {\color{black}Non uniform mechanical strain can also generate unusual electronic and photonic effects \cite{Guinea,Huang,Levy, Cocco,Choi,Lee,Pereira1,Gong,Srivastava,Young,Feng}. Nevertheless, generating well-controlled strain profiles in experiments remains difficult. Typically, strain is applied by mechanically deforming the supporting substrate, either through bending or stretching, which produces in-plane tension or compression, shear strain, and out-of-plane distortions such as ripples or buckling. These strain-induced structural features are commonly characterized using Raman spectroscopy, a powerful and widely used tool for studying strained graphene.}

{\color{black} The coupling between intense electromagnetic fields and graphene is highly non-adiabatic and often irreversible \cite{Higuchi}. The characteristic linear dispersion near the Dirac points, combined with the singular interband coupling, causes a significant  transfer of electrons from the valence to the conduction band, creating a high population of conduction-band electrons that persists even after the pulse ends \cite{Winzer2010, Brida2013}. These interactions can generate ultrafast currents with densities several orders of magnitude higher than those in conventional dielectrics or metals, making graphene an excellent candidate for high-speed optoelectronic applications \cite{Higuchi,Yoshikawa2017}. Moreover, the combination of strong fields with graphene’s two-dimensional nature gives rise to novel phenomena, including {\color{black}Floquet engineering \cite{r1,r2}, light-induced dynamical gaps \cite{r1,r2}, and photon-assisted transport \cite{Wang2013, Mahmood2016,r3}, enabling control over electronic, optical, and topological properties on ultrafast timescales}. Although graphene is a semimetal with zero band gap in the absence of external fields, its electron dynamics under strong electric or laser fields is far from trivial \cite{Winzer2010, Karch2010}. Electrons can traverse the entire Brillouin zone, leading to an effective band offset and the opening of an induced gap of several electronvolts \cite{Oka2009, Sentef2015}. The extraordinary nonlinear response originates from graphene’s unique electronic structure, particularly the singular interband coupling near the Dirac points, which governs the observed irreversibility in electron dynamics \cite{Higuchi}. Recent theoretical and experimental studies have explored a variety of phenomena in graphene under external perturbations. These include {\color{black}laser-induced superlattices \cite{r5}, anisotropic Dirac cone renormalization \cite{r6}, and the combined effects of strain and external fields}. In particular, several works have investigated how strain influences electron transmission through laser barriers \cite{chnafa1}, how transmission behaves in graphene when both laser and magnetic fields are applied simultaneously \cite{chnafa3}, and methods for tuning the energy gap in graphene via laser barriers \cite{chnafa2}. These studies demonstrate the rich interplay between structural deformations, external fields, and light-matter interactions, emphasizing the potential for next-generation devices \cite{Calvo2011, Grushin2014}.}

{\color{black}Motivated by the findings reported in \cite{chnafa1,chnafa2,chnafa3}, we explore the combined impact of an energy gap, a scalar potential, and uniaxial zigzag strain on electron transport in graphene under a laser-modulated barrier. Our model divides the system into three regions, with the central region simultaneously exposed to the laser field, the scalar potential, and the strain. The Dirac equation is solved to obtain the electronic spectrum and wavefunctions for each region. The transmission probabilities are calculated via the transfer matrix method and appropriate boundary conditions. Numerical analysis then reveals how the combined effects of the gap, scalar potential, and zigzag strain influence tunneling through the laser barrier. Because treating all modes numerically is challenging, we restrict our analysis to the first three bands: the central band ($m=0$) and the two adjacent side bands ($m=	\pm1$).  {\color{black}Consequently, we show that the transmissions associated with the three bands $T_0$, $T_{-1}$ and $T_{+1}$ exhibit a strong dependence on strain, gap, scalar potential, and laser parameters. In the absence of strain, the transmission through the central and lower sidebands is generally reduced when either the energy gap or the potential is raised. With the zigzag strain switched on, Fano-type oscillations become prominent, gradually vanishing as the strain increases. The transmission profile increases for low potential values and decreases for high ones. In the upper sideband, the incidence energy shifts the resonance peaks to the right, while enlarging the barrier width generates characteristic oscillatory patterns. Additionally, transmission is enhanced by stronger laser field amplitudes, whereas higher laser frequencies generally reduce it. }}

The structure of this paper is as follows. In Sec. \ref{TFor}, we present the theoretical framework, introducing the Hamiltonian used to investigate the combined impacts of an energy gap, a scalar potential, and uniaxial zigzag strain on electron transport in graphene subjected to a laser-modulated barrier. We then solve the eigenvalue equation and apply Floquet theory to obtain the energy spectrum. These solutions, together with the boundary conditions and the transfer-matrix approach, allow us to determine the transmissions associated with the different energy modes, as described in Sec.  \ref{TFSor}. To further explore, Sec. \ref{TFSor1} offers a numerical analysis and discussion of the transmission behavior for the central band and the first sidebands. Finally, Sec.  \ref{TSFOR2} summarizes our main conclusions.

\section{Theoretical formulation}\label{TFor}	
\begin{figure}[H]\centering
	{\includegraphics[scale=0.55]{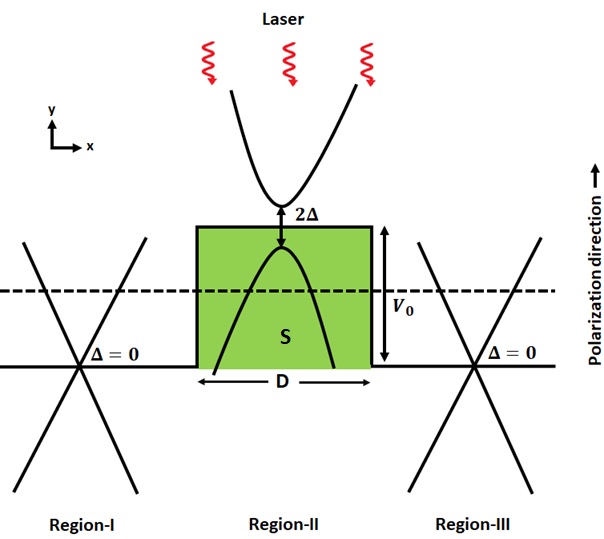}\label{zu}}\\
	{\includegraphics[scale=0.55]{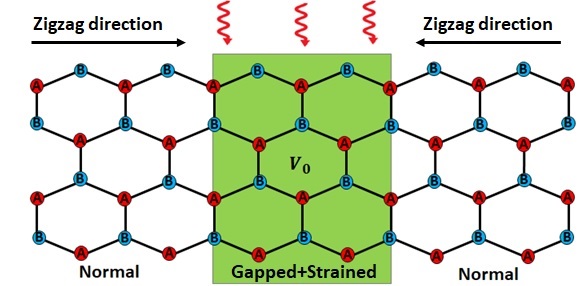}\label{zd}}
	\caption{\color{black}{Illustration of a graphene sheet exposed to a linearly polarized monochromatic laser field, and subjected to a mass term $\Delta$,  mechanical strain along zigzag direction,   and an electrostatic potential $V_0$.}}\label{fz4}
\end{figure}

In this study, we investigate the transport properties of Dirac fermions in graphene sheet divided into three distinct regions. {\color{black}Regions I and III consist of pristine graphene, where charge carriers behave as massless Dirac particles.} In contrast, {\color{black}region II corresponds to a gapped graphene segment subjected to mechanical strain and laser irradiation, which together modify the effective Dirac Hamiltonian}. {\color{black}This central region is additionally exposed to an external electrostatic potential, creating a tunable barrier whose electronic properties differ from those of the pristine sections}. The overall configuration, illustrated schematically in Fig. \ref{fz4}, allows us to explore how the combined effects of a band gap, strain, and a time-dependent laser field influence the dynamics and scattering behavior of Dirac fermions.

For completeness, we comment on the role of the valley degree of freedom. In graphene, the low-energy electronic structure consists of two inequivalent valleys ($K$ and $K'$), and a fully general effective Hamiltonian may include an explicit valley index, as discussed, e.g., in \cite{PhysRevB2023}. In the present work, we employ a single-valley Dirac Hamiltonian, which is justified in the absence of intervalley scattering. This approximation is valid when the system lacks atomically sharp disorder or short-range potentials that could couple the two valleys. Under such conditions, the two valleys are related by time-reversal symmetry and yield identical energy spectra and transport properties. Therefore, the results presented here are valley-independent, up to a trivial degeneracy factor of two. The single-valley Dirac Hamiltonian is
\begin{align}\label{ham2}
	\mathcal{H}=&{v_{x}(S)} {{{\sigma_x}}} \left(p_x-
	\frac{e}{c}{\color{black}{A}_{x}}\right)+{v_{y}(S)}
	{{\color{black}{\sigma_y}}} \left(p_y-
	\frac{e}{c}{\color{black}{A}_{y}}\right)\notag\\
	&+{{\Delta}\sigma_z}{+V\mathbb{I}_2}
\end{align}
where  $\sigma_x$ and $\sigma_y$ are the Pauli matrices acting in the sublattice (pseudospin) space,  $\vec{p}=-i\hbar(\partial_x,\partial_y)$ is the momentum operator, $e$ is the electron charge, $A_x$ and $A_y$ represent the components of the vector potential that couple the Dirac fermion to the applied laser field, $\mathbb{I}_2$ is the $2\times2$ identity matrix in the same space, $\Delta$ is the energy gap. 	
	The  Fermi velocity components $v_x(S)$ and $v_y(S)$  depend on the applied uniaxial strain $S$ along the zigzag direction. This dependence arises from the modification of the nearest-neighbor hopping parameters $t_i$, which are highly sensitive to strain-induced changes in the interatomic bond lengths, as discussed in \cite{Pereira1}. Uner deformation,  the bond lengths become direction-dependent, leading to anistropic hopping amplitudes and, consequnetly, different effective carrier velocities along the $x$- and $y$-directions. Within the low-energy approximation, a linear expansion of the energy dispersion around the Dirac point $K$ gives rise to anisotropic Dirac cones, whose slopes along the $k_x$- and $k_y$-directions define the velocities $v_x(S)$ and $v_y(S)$, given by
\begin{align}\label{xtyr}
&v_{x}(S)=
\frac{\sqrt{3}c}{\hbar}\left(1+ S\right)
\sqrt{t^{2}_{1}-\frac{t^{2}_{3}}{4}}\\
& v_{y}(S)=
\frac{{3}c}{2\hbar}\left(1-\sigma S\right)t_{3}
\end{align}
such that  the hopping energy altered due to mechanical deformation is
\begin{align}
t_{i}=t_{0}e^{-3.37\left({|\xi_{i}|}/{c}-1\right)}	
\end{align}
with  $t_{0}=2.7$ \text{eV}  is the transfer energy, $c=0.165$ \text{nm} denotes the interatomic distance of a covalent bond in the unstrained structure,  the displacement components of deformed graphene sheet along the zigzag direction read as
\begin{align}\label{xrr}
&|\xi_{1}|=|\xi_{2}|=c\left(1+\dfrac{3}{4}S-\dfrac{1}{4}\sigma S\right)\\
& |\xi_{3}|=c\left(1-\sigma S\right)
\end{align}
and $\sigma$ is the Poisson ratio.
In this study,   $v_x(S)$ and $v_y(S)$ are defined within the strip $0 \leq x \leq L$, outside this region we use the unstrained values $v_x (S=0) $ and $v_y(S=0)\rightarrow v_{F}$. 
The electrostatic potential profile along the $x$-axis is written as
\begin{align}\label{HP}
V(x)=\begin{cases}
	V_{0}, & 0<x<D\\
	0,& \text{elsewhere}
\end{cases}
\end{align}
which  can be created in various ways, notably by employing an external electric field or through local chemical doping \cite{wallace,Novoselov,Saito}. Precisely, in the presence of an external electric field, a voltage is applied to electrodes on both sides of the graphene sheet. In contrast, local chemical doping modifies the electronic structure of graphene by introducing impurities or dopant atoms at selected location.
In the dipole approximation, the laser light is represented by the vector potential 
\begin{align}\label{H1f}
\vec{A}=\left(0,A_y\right)=A\cos{\omega t}\ \vec{e}_y
\end{align}
where the unit vector $\vec{e}_y$   along the $y$-axis denotes the direction of laser polarization, $\omega$ is the frequency of a wave. The corresponding laser field is
\begin{eqnarray}\label{xt7}
{\vec{F}=-\dfrac{1}{c}\dfrac{\partial}{\partial t}\vec{A}=F\sin{\omega t}} \ \vec{e}_y
\end{eqnarray}
with {$F=\dfrac{\omega}{c}{A}$ represents its amplitude.

From the above consideration, we can express the eigenvalue equation $	\mathcal{H}\Phi_\text{j}(x,y,t)=\mathcal{E}\Phi_\text{j}(x,y,t)$ associated with   regions \text{j=1,2,3} as
\begin{widetext}
	\begin{align}\label{H2}
		\left\{v_{x}(S)\left[{\sigma}_xp_x+\dfrac{v_{y}(S)}{v_{x}(S)}{\sigma}_y\left({p}_y-\dfrac{eF}{\omega}\right)\right]+{{\Delta}\sigma_z}{+V\mathbb{I}_2}\right\}\Phi_{\text{j}}(x,y,t)=i\hbar\partial_t \Phi_{\text{j}}(x,y,t).
	\end{align}
\end{widetext}
By including  the laser field, both the electron wavefunction and the band structure of graphene undergo modifications, which are addressed within the framework of the Floquet approximation. Given that the laser pulse is modeled as continuous wave, the solution to (\ref{H2}) adopts the following configuration
\begin{widetext}	\begin{align}\Phi_\text{j}(x,y,t)
		=\begin{pmatrix}\Phi_{\text{A},\text{j}}(x,y,t)\\\Phi_{\text{B},\text{j}}(x,y,t)\end{pmatrix}=e^{-i\mathcal{E}t}\xi_\text{j}(t)\phi_\text{j}(x,y).
\end{align} \end{widetext} This leads to the coupled equations
\begin{widetext}
	\begin{align}\label{re}	
		&\left({\Delta +V_0}\right)\xi_\text{j}(t)\phi_{\text{A},\text{j}}(x,y)+  \left[-iv_{x}(S){\partial_x}-v_{y}(S){\partial_y}+i{\dfrac{v_{y}(S)F}{\omega}}\cos{\omega t}\right]\xi_\text{j}(t)\phi_{\text{B},\text{j}}(x,y) =i\partial_t\xi_\text{j}(t) \phi_{\text{A},\text{j}}(x,y)\\\label{re1}
		&\left[	-iv_{x}(S){\partial_x}+{v_{y}(S)\partial_y}-i{\dfrac{v_{y}(S)F}{\omega}}\cos{\omega t}\right]\xi_\text{j}(t)\phi_{\text{A},\text{j}}(x,y) + \left[{-\Delta +V_0}\right]\xi_\text{j}(t)\phi_{\text{B},\text{j}}(x,y)
		=i\partial_t \xi_\text{j}(t) \phi_{\text{B},\text{j}}(x,y)
	\end{align}
\end{widetext}
where the dimensionless parameters are specified  $F\equiv{F}/{{F_0}}$, $\textbf{r}\equiv{\textbf{r}}/{{D_0}}$, ${D\equiv{D}/{{D_0}}}$, $v_{x}(S)=v_{x}(S)/v_F$, $v_{y}(S)=v_{y}(S)/v_F$, $\varepsilon\equiv{\varepsilon}/{\varepsilon_{0}}$, ${\Delta}\equiv{\Delta}/{\varepsilon_{0}}$, ${ V_0}\equiv{V_0}/\varepsilon_{0}$, $t\equiv{t}/{t_{0}}$, $\omega\equiv{\omega}/{\omega_{0}}$, with scales:  length $D_{0}=\sqrt{\dfrac{\hbar v_{F}}{eF_{0}}}$, time  $t_{0}=\dfrac{{D_0}}{v_{F}}$, energy  $\varepsilon_{0}=\dfrac{\hbar {\tiny }v_{F}}{{D_0}}$, frequency  $\omega_{0}=\dfrac{v_{F}}{{D_0}}$ and  electric field  $F_0=\dfrac{\varepsilon_{0}}{e{D_0}}$.
%
{Yet, these two equations alone do not allow for the determination of three unknown functions $\phi_{\text{A},\text{j}}$, $\phi_{\text{B},\text{j}}$ and $\xi_\text{j}(t)$. Consequently, an approximation method must be employed. Here, we use an iterative technique to resolve these variables. Initially, we make the assumption that $\phi_{\text{A},\text{j}}(x,y)$, $\phi_{\text{B},\text{j}}(x,y)$ adhere to the coupled differential equations within the barrier region without the influence of the laser.  Based on this assumption, (\ref{re}) and (\ref{re1}) can be simplified to 
	\begin{align}\label{re2}	
		{\dfrac{v_{y}(S)F}{\omega}}\cos{\omega t}\xi_\text{j}(t)\phi_{\text{B},\text{j}}(x,y) &=\partial_t\xi_\text{j}(t) \phi_{\text{A},\text{j}}(x,y)\\\label{re3}
		-{\dfrac{v_{y}(S)F}{\omega}}\cos{\omega t}\xi_\text{j}(t)\phi_{\text{A},\text{j}}(x,y) 
		&=\partial_t \xi_\text{j}(t) \phi_{\text{B},\text{j}}(x,y).
	\end{align}
	Since $\phi_{\text{A},\text{j}}(x,y)$ and $\phi_{\text{B},\text{j}}(x,y)$ are independent of time, we have $\partial_t\phi_{\text{A},\text{j}}(x,y)=\partial_t\phi_{\text{B},\text{j}}(x,y)=0$. By differentiating (\ref{re2}), we obtain
	\begin{widetext}
		\begin{align}\label{rep}	
			\partial^{2}_{t} \xi_\text{j}(t) \phi_{\text{A},\text{j}}(x,y)
			= \frac{v_{y}(S)F}{\omega} \left[ -\omega \sin(\omega t)\xi_\text{j}(t) \phi_{\text{B},\text{j}}(x,y)
			+ \cos(\omega t) \partial_t\xi_\text{j}(t) \phi_{\text{B},\text{j}}(x,y) \right].
		\end{align}
	\end{widetext}
	(\ref{re3}) allows us to write (\ref{rep}) as
	\begin{widetext}
		\begin{align}\label{repm}	
			\partial^{2}_{t} \xi_\text{j}(t) \phi_{\text{A},\text{j}}(x,y)
			= \frac{v_{y}(S)F}{\omega} \left[ -\omega \sin(\omega t)\xi_\text{j}(t) \phi_{\text{B},\text{j}}(x,y)
			+ \cos(\omega t) \left(-{\dfrac{v_{y}(S)F}{\omega}}\cos{\omega t}\xi_\text{j}(t)\phi_{\text{A},\text{j}}(x,y) \right) \right].
		\end{align}	
	\end{widetext}
	Using (\ref{re2}) to express $\phi_{\text{B},\text{j}}(x,y)$ in terms of $\phi_{\text{A},\text{j}}(x,y)$
	\begin{align}\label{rmp}	
		\phi_{\text{B},\text{j}}(x,y)=\dfrac{ \partial_t\xi_\text{j}(t)}{{\dfrac{v_{y}(S)F}{\omega}} \cos(\omega t) \xi_\text{j}(t)}	\phi_{\text{A},\text{j}}(x,y)
	\end{align}
	leading to the following second order differential equation 
	\begin{align}\label{gg1}
		\partial^{2}_t\xi_\text{j}(t)+\omega\tan\omega{t} \partial_t\xi_\text{j}(t)+\dfrac{{v_{y}^{2}(S)F}^{2}}{{\omega}^{2}}\cos^{2}\omega{t}\xi_\text{j}(t)=0
	\end{align}
	which can be solved to find
	\begin{align}\label{g1}
		\xi_\text{j}(t)=e^{-i\frac{v_{y}F}{\omega^{2}}\sin{\omega t}}=\sum^{
			+\infty}_{l=-\infty}J_{l}\left(\nu\right)e^{-i l\omega t}
	\end{align}
	where $J_{l}$ denotes the first-order Bessel function and $\nu=v_{y}F/\omega^{2}$ is  its argument. Bringing together all components to formulate the spinor of $\mathcal{H}$ as
	\begin{align}\label{1}
		\Phi_\text{j}(x,y,t)=\phi_\text{j}(x,y)\sum^{+\infty}_{l=-\infty}J_{l}\left(\nu\right)e^{-i \left(\mathcal{\varepsilon}+l\omega \right)t}
	\end{align}
	
	It is clear that determining $\phi_\text{j}(x,y)$ is essential to obtain a complete solution. The wave functions within {regions I $(x<0)$ and III $(x>D)$}, devoid of any potential influence (whether from gate voltage or laser, i.e., {$V=0$ and $F=0$}) with the energy gap being inactive ({$\Delta=0$}) can be expressed as
	\begin{widetext}
		\begin{align}
			&	\Phi_{\text{I}}(x,y,t)={e^{ik_{y}y}}\sum^{+\infty}_{m,l=-\infty}\left[\delta_{m0}\begin{pmatrix}
				1\\
				\aleph_m
			\end{pmatrix}e^{i{k_{m}}x}
			+ r_{m}  \begin{pmatrix}
				1\\
				-\dfrac{1}{\aleph_m}
			\end{pmatrix}e^{-i{k_{m}}x}\right]\delta_{lm}e^{-i\left(\mathcal{\varepsilon}+l\omega\right)t}
			\\
			&
			\Phi_{\text{III}}(x,y,t)={e^{ik_{y}y}}\sum^{+\infty}_{m,l=-\infty}\left[t_{m}\begin{pmatrix}
				1\\
				\aleph_m
			\end{pmatrix}e^{i{k_{m}}x}
			+ \Xi_{m}  \begin{pmatrix}
				1\\
				-\dfrac{1}{\aleph_m}
			\end{pmatrix}e^{-i{k_{m}}x}\right]\delta_{lm}e^{-i\left(\mathcal{\varepsilon}+l\omega\right)t}
		\end{align}
	\end{widetext}
	where $\Xi_m$ is the vector of zero magnitude and $r_{m}$, $t_{m}$ are the reflection and transmission amplitudes, respectively. The complex number and the angle  outside the barrier  are defined as
	\begin{align}
		&	{\aleph_m}=s_m\dfrac{k_{m}+ik_y}{\sqrt{k_{m}^{2}+k^{2}_{y}}}=s_m e^{i\theta_m}\\
		& \theta_m=\tan^{-1}\left(\frac{k_y}{k_{m}}\right)
	\end{align}
	$s_m$ represents the sign function of $(\varepsilon+m\omega)$, where $k_{x,m}$ is denoted by $k_m$. The energies corresponding to each state, along with their $x$-component wave vectors, are detailed as
	\begin{align}\label{a1}
		&	{\mathcal{\varepsilon}+m\omega=s_m\sqrt{k_{m}^{2}+k^{2}_{y}}}\\
		&
		k_{m}=\sqrt{\left(\mathcal{\varepsilon}+m\omega\right)^{2}-k^{2}_{y}}.
	\end{align}
	For region-\text{II} ($0< x< D$)}, {once the effects of the sidebands, the laser field ($F\neq 0$), the potential vector ($V\neq0$) and the energy gap ($\Delta\neq 0$) are taken into account, the complete solution for the pseudospin state within the barrier can be formulated as 
	\begin{widetext}
		\begin{align}\label{AZ}
			\Phi_{\text{II}}(x,y,t)&={e^{ik_{y}y}}\sum^{
				+\infty}_{m,l=-\infty}\left[\alpha_{m}\begin{pmatrix}
				\zeta^{+}_{m}\\
				\zeta^{-}_{m}\eta'_{m}
			\end{pmatrix}e^{i{q_{m}}x}+ \beta_{m} \begin{pmatrix}
				\zeta^{+}_{m}\\
				-\dfrac{\zeta^{-}_{m}}{\eta'_m}
			\end{pmatrix}e^{-i{q_{m}}x}\right]J_{l-m}\left(\nu\right)e^{-i\left(\mathcal{\varepsilon}+l\omega\right)t}
		\end{align} 
	\end{widetext}
	wherein we have established
	\begin{align}
		\zeta^{\pm}_{m}=\left(1\pm\dfrac{s'_m\Delta}{\sqrt{v^{2}_{x}(S)q_{m}^{2}+(v_{y}(S)k_{y}-m\omega)^{2}+\Delta^{2}}}\right)^{\frac{1}{2}}	\end{align}
	together with the parameters
	\begin{align}\label{mz7}
		&\eta'_m=s'_m\dfrac{v_{x}(S)q_{m}+i\left(v_{y}(S)k_{y}-m\omega\right)}{\sqrt{{v^{2}_{x}(S)q_{m}^{2}}+(v_{y}(S)k_{y}-m\omega)^{2}}}=s'_m e^{i\theta'_m}\\
		& \theta'_m=\tan^{-1}\left(\frac{v_{y}(S)k_y-m\omega}{v_{x}(S)q_{m}}\right)
	\end{align}
	where {$s'_m=\text{sgn}(\varepsilon+m\omega-V_0)$}, $q_{x,m}=q_m $, $\alpha_{m}$ and $\beta_{m}$ are two constants. 
	The eigenvalues that correspond to (\ref{AZ}) read as
	\begin{align}\label{m7}
		&	{\mathcal{\varepsilon}+m\omega=V_0+s'_m\sqrt{{v^{2}_{x}(S)q_{m}^{2}+(v_{y}(S)k_{y}-m\omega)^{2}{+\Delta^{2}}}}}
	\end{align}
	yields the wave vector
	\begin{align}
		&\label{A1}
		q_{m}=\sqrt{\left(\frac{\mathcal{\varepsilon}+m\omega-V_0}{v_{x}(S)}\right)^{2}-\left(\frac{v_{y}(S)k_{y}-m\omega}{v_{x}(S)}\right)^{2}-\frac{\Delta^{2}}{v^{2}_{x}(S)}}.
	\end{align}
	Considering the corresponding eigenvalues and wave vectors obtained from the relevant equations, we can formulate expressions for the transmission coefficients associated with each energy mode.  By analyzing these transmission probabilities across different energy modes, we obtain a more comprehensive understanding of the transmission behavior of the system, which is essential for understanding its operation and physical properties under various conditions. 
	

	\section{Transmission Chnannels}\label{TFSor}
	
	In studying the transmission probabilities in gapped graphene across a laser-assisted barrier, our initial step involves establishing the continuity of eigenspinors at the interfaces $x=0,{D}$, i.e., ${\Phi_{\text{I}}(0,y,t)}={\Phi_{\text{II}}(0,y,t)}$ and ${\Phi_{\text{II}}({D},y,t)}={\Phi_{\text{III}}({D},y,t)}$. Next, considering that $e^{il{\omega} t}$ as an orthogonal basis, we then evaluate the conditions at $x=0$ on this basis, which leads to the following system of equations
	\begin{align}
		&\delta_{l0}+r_{l} =\sum^{+\infty}_{m=-\infty}\zeta^{+}_{m}
		\left[\alpha_{m}+\beta_{m}\right]
		J_{l-m}{{\left(\nu\right)}} \label{eqx01}\\
		&\delta_{l0}{\aleph_l}-\dfrac{{r_{l}}}{\aleph_l}=
		\sum^{+\infty}_{m=-\infty}{{\color{red}}\zeta^{-}_{m}\left[\alpha_{m}{\eta'_m}-\frac{\beta_{m}}{{{\eta'_m}}}\right]}
		J_{l-m}{{\left(\nu\right)}} \label{eqx02}	.
	\end{align}
	Similarly, applying the continuity conditions at $x=D$ and projecting onto the same orthogonal basis, we obtain
	\begin{widetext}
		\begin{align}
				&t_{l}{e^{i{k_{l}}{D}}}+
				{\Xi_l}{e^{-i{k_{l}}{{D}}}} =\sum^{+\infty}_{m=-\infty} \zeta^{+}_{m}\left[\alpha_{m}{e^{i{q_{m}}{D}}}+\beta_{m}{e^{-i{q_{m}}{D}}}\right]J_{l-m} {{\left(\nu\right)}} \label{eqxd1}\\
				&	t_{l}{\aleph_l}{e^{i{k_{l}}{{D}}}}-\dfrac{{\Xi_l}}{\aleph_l}{e^{-i{k_{l}}{D}}} 
				=\sum^{+\infty}_{m=-\infty}{\color{red}}\zeta^{-}_{m}\left[\alpha_{m}{\eta'_{m}}{e^{i{\color{black}q_{m}}{D}}}-\frac{\beta_{m}}{{\eta'_{m}}}{e^{-i{q_{m}}{D}}}\right]J_{l-m} {{\left(\nu\right)}}. \label{eqxd2}
			\end{align}	
		\end{widetext}
		These equations form a complete set for determining the reflection $r_l$ and transmission  $t_l$ amplitudes 
		for each sideband $l$.} They can be conveniently written in a compact matrix form as
	\begin{align}\label{ar1}
		\begin{pmatrix}
			{\delta_{0m}}\\
			{r_m}
		\end{pmatrix}=\begin{pmatrix}
			{\mathbb{C}_{11}}   &{\mathbb{C}_{12}}\\
			{\mathbb{C}_{21}}&{\mathbb{C}_{22}}
		\end{pmatrix}\begin{pmatrix}
			t_m\\
			\Xi{m}
		\end{pmatrix}=\mathbb{C}\begin{pmatrix}
			t_m\\
			\Xi_{m}
		\end{pmatrix}
	\end{align}
	where the matrix $\mathbb{C}$ is given by
	\begin{widetext}
		\begin{align}
			{\mathbb C}={\begin{pmatrix}
					{\mathbb 1}& {\mathbb 1} \\
					{{\mathbb U^{+}}} &{{\mathbb U^{-}}} \\
				\end{pmatrix}^{-1}}\cdot
			\begin{pmatrix}
				{{\mathbb W^{+}(0)}} & {{\mathbb W^{-}(0)}} \\
				{{\mathbb Y^{+}(0)}} & {{\mathbb Y^{-}(0)}}
			\end{pmatrix}\cdot \begin{pmatrix}
				{{\mathbb W^{+}(D)}} & {{\mathbb W^{-}(D)}} \\
				{{\mathbb Y^{+}(D)}} & {{\mathbb Y^{-}(D)}}
			\end{pmatrix}^{-1}\cdot\begin{pmatrix}
				{\mathbb 1}& {\mathbb 1} \\
				{{\mathbb U^{+}}} &{{\mathbb U^{-}}} \\
			\end{pmatrix}\cdot\begin{pmatrix}
				{{\mathbb Z^{+}(D)}}& {\mathbb O} \\
				{{\mathbb O}} &{{\mathbb Z^{-}(D)}} 
			\end{pmatrix}
		\end{align}
	\end{widetext}
	such that ${\mathbb O}$ is the null matrix,  ${\mathbb I}$ is the unit matrix, and the  matrix elements are defined as
	\begin{align} \label{eqn1}
		&	\left({{\mathbb
				U^{\pm}}}\right)_{lm}=\pm\left({\aleph_{l}}\right)^{\pm
			1}\delta_{lm}\\
		&
		\left({{\mathbb
				W^{\pm}(x)}}\right)_{lm}=\zeta^{+}_{m}e^{\pm iq_{m}{x}}J_{l-m}{{\left(\nu\right)}}, \\ &\left({{\mathbb Y^{\pm}(x)}}\right)_{lm}=\pm \zeta^{-}_{m} ({\eta'_{m}})^{\pm 1}e^{\pm iq_{m}{x}}
		J_{l-m}{{\left(\nu\right)}}\\
		&	\left({\mathbb Z^{\pm}(D)}\right)_{lm}=e^{\pm i{k_{l}}{D}}\delta_{lm}.
	\end{align}
	This matrix formalism allows us to systematically relate the incident, reflected, and transmitted amplitudes through a single operator $\mathbb{C}$. By inverting the submatrix $\mathbb{C}_{11}$, we can explicitly obtain the transmission  amplitude for each sideband 
	\begin{align}\label{n1}
		&t_{m}={\mathbb{C}_{11}^{-1}} \cdot\delta_{0m}
	\end{align}
	This approach provides a compact and efficient method for calculating the full scattering amplitudes in the presence of multiple sidebands, making it particularly suitable for numerical implementation. Remember that the value of $m$ ranges from $-N$ to $N$  \cite{k1}, then we can write \eqref{n1} as
	\begin{align}
		\begin{pmatrix}
			t_N  \\
			\cdot\\
			\cdot\\
			t_{-1}\\
			t_0	\\
			t_1\\
			\cdot\\
			\cdot\\
			t_{N}
		\end{pmatrix}={\mathbb{C}_{11}^{-1}}\cdot\begin{pmatrix}
			0  \\
			\cdot\\
			\cdot\\
			0\\
			1	\\
			0\\
			\cdot\\
			\cdot\\
			0
		\end{pmatrix}.
	\end{align} 
	For simplicity, we consider the first three bands numerically: the central band associated with the energy ${\mathcal{\varepsilon}}$, and the two adjacent side bands associated with the energies ${\mathcal{\varepsilon}+\omega}$ and ${\mathcal{\varepsilon}-\omega}$ \cite{k1,chnafa1, chnafa2,El Aitouni,AZ}. This results in
	\begin{align}\label{A3}
		t_{-1}={\mathbb{C}_{11}^{-1}}[1,2], \quad 
		t_{0}={\mathbb{C}_{11}^{-1}}[2,2],\quad 
		t_{1}={\mathbb{C}_{11}^{-1}}[3,2]
	\end{align}
	
	To determine the transmission probabilities $T_{m}$, we introduce the current density $\boldsymbol{J}$. Using  the continuity equation, we obtain
	%
	\begin{align}
		\boldsymbol{J}=\psi^\dagger(x,y,t) \sigma_{x}\psi(x,y,t).
	\end{align}
	This gives rise to the transmitted current density ${\boldsymbol{J}_{\textbf{tran},m}}$ and the incident current density ${\boldsymbol{J}_{\textbf{inc}, 0}}$
	\begin{align}
		&{\boldsymbol{J}_{\textbf{inc}, 0}}=\aleph^{\ast}_{0}+\aleph_{0}\\
		&
		{\boldsymbol{J}_{\textbf{tran},m}}=t^{\ast}_mt_m\left(\aleph^{\ast}_{m}+\aleph_{m}\right)
	\end{align}
	They connect the microscopic wavefunction amplitudes to measurable transport properties, guaranteeing current conservation across the different Floquet sidebands.
	Consequently, from the relation $T_{m}=\left|\frac{{\boldsymbol{J}_{\textbf{tran},m}}}{{\boldsymbol{J}_{\textbf{inc}, 0}}}\right|$, we get
	%
	%
	\begin{align}\label{Am} 
		&T_{m}=\frac{s_m}{s_0}\frac{\cos{\theta_m}}{\cos{\theta_0}}\lvert t_m\rvert^{2}\\
		& {\cos{\theta_m}}=\dfrac{k_{m}}{\sqrt{k_{m}^{2}+k^{2}_{y}}}.
	\end{align}
	These results will be numerically analyzed to highlight how deformation along the zigzag direction, potential, and the gap influence the control of electronic transport in graphene. 
	
	\section{Results and discussions}\label{TFSor1}
	
	Fig. \ref{h} highlights the combined influence of the strain magnitude $S$ and the energy gap $\Delta$ on the transmission probability for central band $T_0$ versus the barrier height $V_0$, under the conditions $k_y=2$, $D=1.2$, $F=0.6$, $\omega=1$,  $\varepsilon=15$, $\Delta=1$ (blue line), $\Delta=3$ (green line), $\Delta=5$ (red line). In Fig. \ref{h}(a), in the absence of strain ($S=0.0$), we observe that for $\Delta=1$, the transmission displays three distinct regions,  each reflecting a different transport behavior. In the first region $\{0\leq V_0\leq13\}$, $T_0$ shows oscillatory behavior and generally decreases  as $V_0$ increases as noted in \cite{k1,delta,AZ}. These oscillations arises from quantum interference phenomena. Electrons in graphene behave like wave: when they encounter a potential barrier, part of the wave is reflected and part is transmitted.  The interaction between the reflected and transmitted waves generates transmission maxima and minima, which explains the observed oscillatory behavior. Moreover, the drop in $T_0$ occurs because the energy difference between the electron states and the barrier increases. This makes tunneling less likely, leading to a general decrease in transmission. Subsequently, they disappear in the region $\{13\leq V_0\leq17\}$, inside the barrier becomes imaginary, indicating that the electrons no longer have sufficient energy to pass through the barrier. Instead of propagating, the wavefunction decays exponentially within the barrier, giving rise to what are known as evanescent states.   When the potential $V_0$ exceeds the value $13$, the transmission increases and then tends to stabilize at a constant level ($T_0=0.8$). This stabilization can be explained by the fact that interference effects become negligible and that part of the electrons pass through the barrier via quasi-propagating or resonant states. As long as 
	$\Delta$ increases, the transmission decreases showing a behavior similar to that observed in \cite{delta,AZ}, the number of oscillations increases, and the transmission width in the interval $\{9.7\leq V_0\leq21\}$ becomes larger. 
	The observed features result from the competition between gap-induced suppression of electron transport and resonance–interference mechanisms that permit transmission at selected energies. 
	Conversely, when strain is applied along the zigzag direction ($S=0.2$), the number of transmission peaks increases and the transmission profile is significantly modified. It rises for small values of $V_0$ and decreases for larger ones, as shown in Fig.~\ref{h}(b). This behavior can be understood as a consequence of the strain-induced modification of graphene electronic band structure, which gives rise to additional resonance states. These results demonstrate that the combined effects of an energy gap and zigzag strain strongly influence the transmission in the central band.
	
	\begin{figure}[H]\centering
		\subfloat[]{\includegraphics[width=0.48\linewidth, height=0.2\textheight]{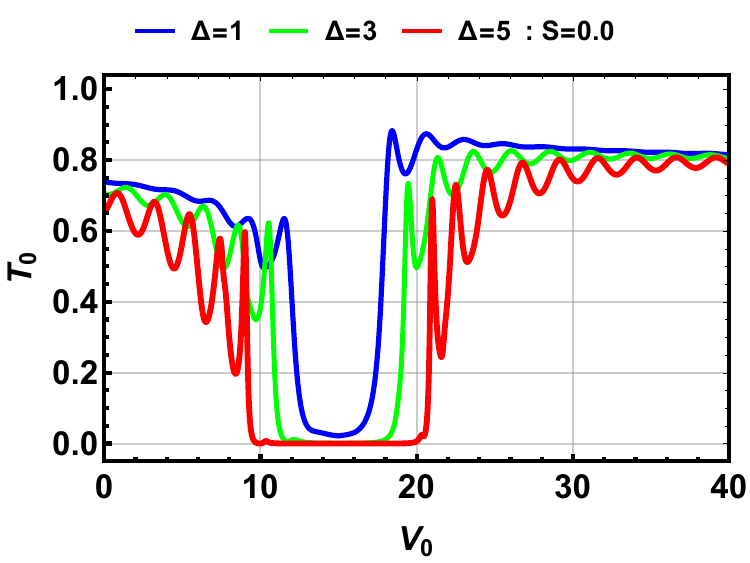}\label{fy}}	\subfloat[]{\includegraphics[width=0.48\linewidth, height=0.2\textheight]{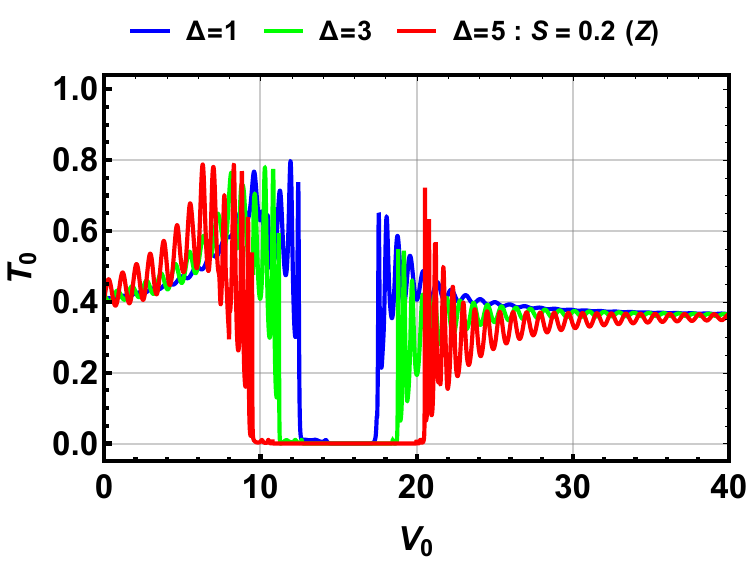}\label{figu}}	
		\caption{Transmission probability for  central band $T_0$  versus the  barrier height $V_0$ for $k_y=2$, $D=1.2$, $F=0.6$, $\omega=1$, $\varepsilon=15$, $\Delta=1$ (blue line), $\Delta=3$ (green line), $\Delta=5$ (red line). (a):  Without strain $S=0.0$,  (b){:}  for zigzag direction $S=0.2$.}\label{h}
	\end{figure}
	
	Fig. \ref{Nh}  shows density plot of $T_0$ versus the  barrier height $V_0$ and the energy gap  $\Delta$ for $k_y=2$, $D=1.2$, $F=0.6$, $\omega=1$, $\varepsilon=15$,
	and $S=0.0, 0.2 $.
	%
	For $S=0.0$ in Fig. \ref{Nh}(a), we notice that $T_0$ exhibits pronounced peaks at $\varepsilon<V_0$ and $\varepsilon>V_0$.  As $\Delta$ increases, a clear transmission gap begins to emerge: the values of $T_0$ decrease sharply around an intermediate range of $V_0$, as displaying a behavior comparable to that found in \cite{density}. This expansion of the transmission gap becomes more significant for larger $\Delta$, confirming the important filtering role played by the energy gap. Once the strain is switched on to the value $S= 0.2 $, as shown in  Fig.~\ref{Nh}(b),  it becomes evident that  $T_0$ exhibits the same behavior as in Fig. \ref{Nh}(a), except that the presence of $S$ introduces additional resonance channels. The resonant features are more visible for small values of $\Delta$ and gradually diminish as the gap increases, indicating that the strain-induced resonances are highly sensitive to the magnitude of the band gap. For sufficiently large $\Delta$, the oscillations of $T_0$ are strongly suppressed and eventually disappear, leaving a region dominated by low  $T_0$. {\color{black}These results show that the interaction between the energy gap and zigzag strain not only governs the appearance and disappearance of transmission resonances but also plays a key role in shaping the overall profile of $T_0(V_0,\Delta)$.} This interplay provides a tunable mechanism for manipulating electron transport in strained gapped graphene under laser irradiation.
	
	\begin{figure}[H]\centering
		\subfloat[]{\includegraphics[width=0.5\linewidth, height=0.16\textheight]{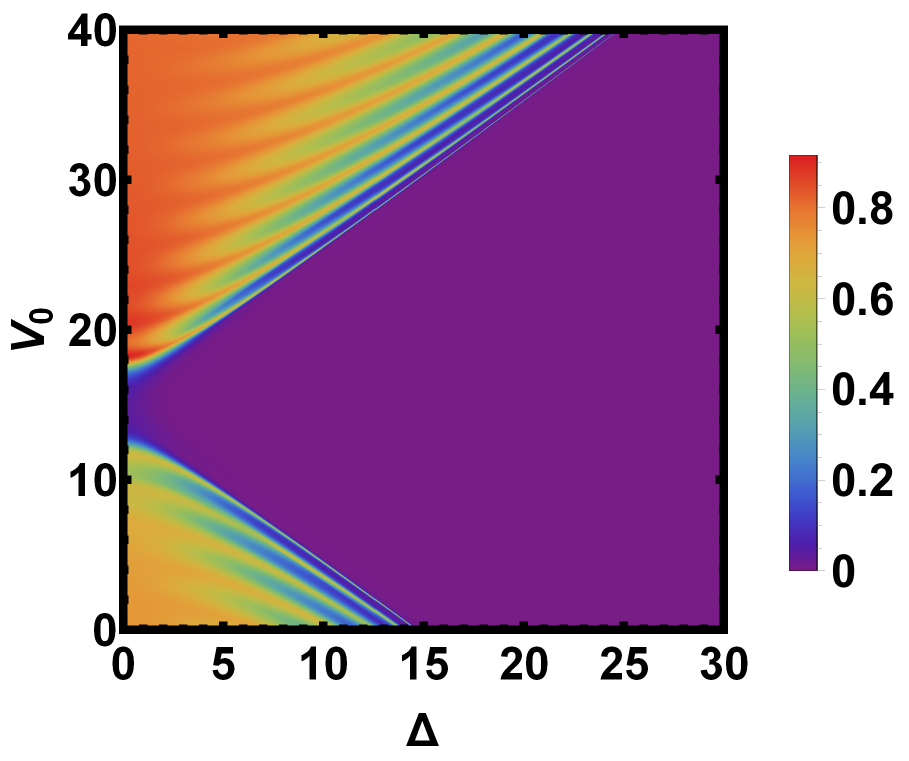}\label{fiy}}	\subfloat[]{\includegraphics[width=0.5\linewidth, height=0.16	\textheight]{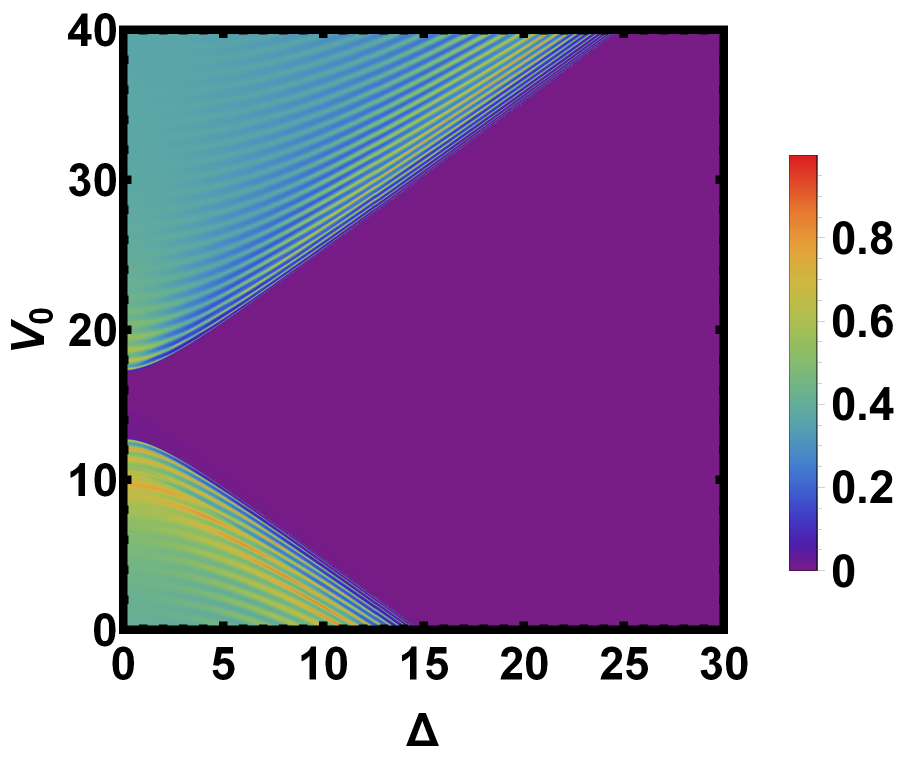}\label{fig}}	
		\caption{Density plot of transmission probability for  central band $T_0$  versus the  barrier height $V_0$ and the energy gap  $\Delta$ for $k_y=2$, $D=1.2$, $F=0.6$, $\omega=1$, $\varepsilon=15$. (a){:} Without strain $S=0.0$, (b){:} for zigzag direction $S=0.2$. }\label{Nh}
	\end{figure}
	
	Fig. \ref{m1} presents the transmission probability for central band $T_{0}$ versus the zigzag strain $S$. In Fig. \ref{m1}\text{{(a)}}, we consider three values of the potential height, $V_0=2$ (blue line), $V_0=4$ (green line), and $V_0=6$ (red line), while the energy gap is fixed at $\Delta=3$.   
	We observe that $T_0$ gradually decreases as  $S$ increases, while exhibiting resonant oscillations whose amplitude depends on the applied potential $V_0$.  For $V_0=2$, the resonances remain pronounced with maxima exceeding $T_0\simeq0.95$  at low strain, whereas for $V_0=4$ their amplitude decreases significantly. In contrast, when $V_0=6$, the transmission is strongly reduced over the whole strain range and the resonances become very weak. Moreover, at larger strain values, the transmission tends toward nearly zero, in agreement with \cite{M1,D2}. Physically, this suppression of transport arises because the zigzag strain strongly alters the band structure of graphene, which  shifts the Dirac cones and enhancing the sublattice asymmetry. These effects effectively enlarge the energy gap and progressively suppress the resonance conditions responsible for finite transmission. As a result, when the strain becomes strong, electrons can no longer find compatible propagating states, tunneling breaks down, and the transport channels are completely blocked, regardless of the value of $V_0$. In Fig.~\ref{m1}\text{{(b)}}, the potential is fixed at $V_0=4$, while the band gap is varied, with $\Delta=1$ (blue line), $\Delta=3$ (green line), $\Delta=5$ (red line). The transmission exhibits the same   behavior as in Fig.~\ref{m1}\text{{(a)}} except that the number of oscillations increases, especially for $\Delta=5$, and their amplitude  becomes significantly larger. 
	This indicates a more pronounced modification of the band structure. 
	
	\begin{figure}[H]\centering
		\subfloat[]{\includegraphics[width=0.5\linewidth, height=0.21\textheight]{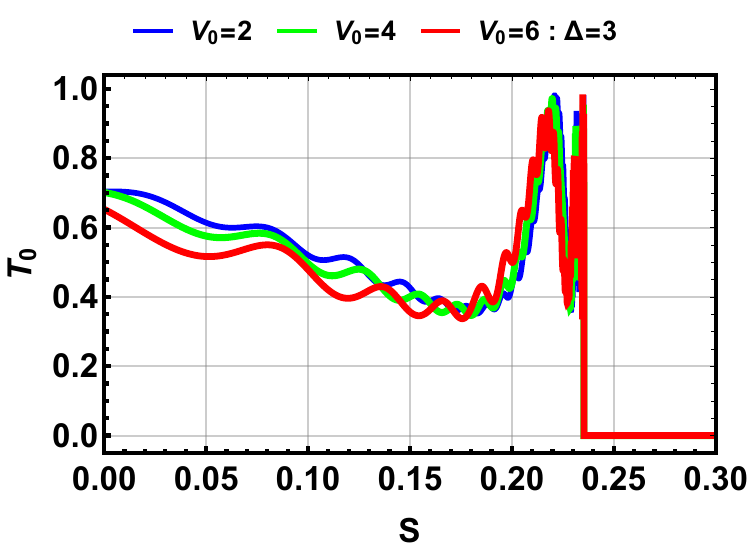}\label{y}}	\subfloat[]{\includegraphics[width=0.5\linewidth, height=0.21\textheight]{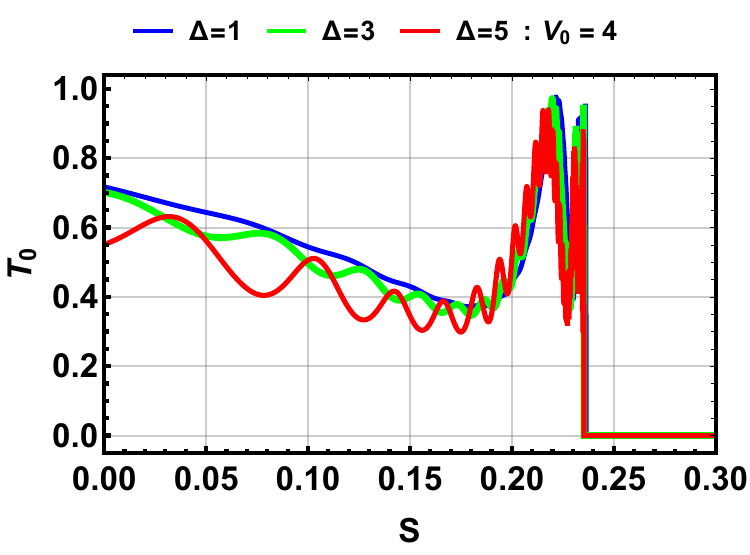}\label{Th}}	
		\caption{{Transmission probability for  central band $T_0$  versus the zigzag strain $S$ for $k_y=2$, $D=1.2$, $F=0.6$, $\omega=1$,  $\varepsilon=15$. \text{{(a)}}\color{black}{:} $V_0=2$ (blue line), $V_0=4$ (green line), $V_0=6$ (red line), $\Delta=3$. \text{{(b)}}\color{black}{:} $\Delta=1$ (blue line), $\Delta=3$ (green line), $\Delta=5$ (red line), $V_0=4$.} }\label{m1}
	\end{figure}

	\begin{figure}[H]\centering
		\subfloat[]{\includegraphics[width=0.5\linewidth, height=0.21\textheight]{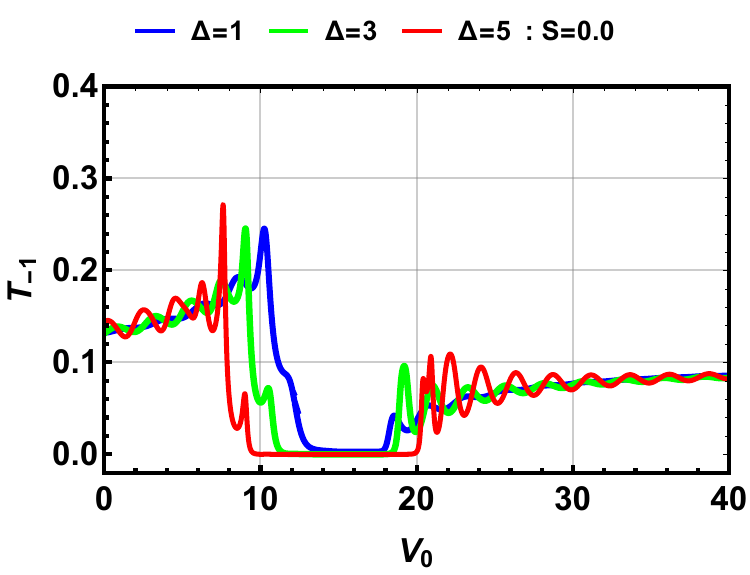}\label{fig1}}	\subfloat[]{\includegraphics[width=0.5\linewidth, height=0.21\textheight]{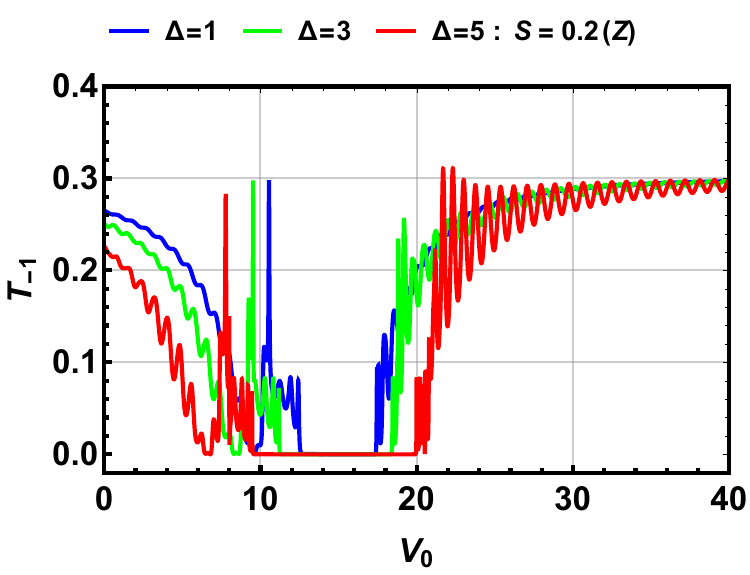}\label{fig2}}
		\caption{{Transmission probability for single photon emission $T_{-1}$ versus the barrier height $V_0$ for $k_y=2$, $D=1.2$, $F=0.6$, $\omega=1$, $\varepsilon=15$, $\Delta=1$ (blue line), $\Delta=3$ (green line), $\Delta=5$ (red line). \text{{(a)}}\color{black}{:}  Without strain $S=0.0$, \text{{(b)}}\color{black}{:}  for zigzag direction $S=0,2$.}}\label{Nfm}
	\end{figure}
	In Fig.~\ref{Nfm}, we plot the transmission probability for single photon emission $T_{-1}$ versus the barrier height $V_0$ for $k_y=2$, $D=1.2$, $F=0.6$, $\omega=1$, $\varepsilon=15$, $\Delta=1$ (blue line), $\Delta=3$ (green line), $\Delta=5$ (red line), and  $S=0.0, 0.2$. Regarding the case $S=0.0$ in  Fig.~\ref{Nfm}(a), the transmission profile of the first band shifts to the left as $V_0$ rises, and more oscillations appear when $\Delta$ becomes larger as seen in \cite{chnafa2,delta,AZ}. This behavior indicates that increasing the potential shifts resonant transmission to lower energies, while a larger gap increases the interaction between electronic states, creating additional resonances. In Fig.~\ref{Nfm}(b), when a zigzag strain ($S=0.2$) is applied, the number of oscillations in the transmission increases significantly. We also observe the emergence of peaks in the range $8<V_0<11$, corresponding to energies where electrons can pass through the barrier more easily. However, the maxima of these peaks decrease as the potential $V_0$ increases, showing that the barrier becomes more effective at blocking electrons and reduces the transmission at resonant energies. 
	Moreover, for all three values of $\Delta$, the transmission increases rapidly with $V_0$. In contrast to Fig.~\ref{Nfm}\text{(a)}, where the transmission remains nearly constant around $T_{-1}\simeq 0.09$, this behavior indicates that increasing the potential significantly modifies the phase conditions and electronic coupling within the barrier. As a result, electron transmission is enhanced and additional transport channels become accessible. These findings highlight the combined roles of deformation, electrostatic potential, and the energy gap in controlling electron transport in graphene.
	
	\begin{figure}[H]\centering
		\subfloat[]{\includegraphics[width=0.5\linewidth, height=0.165\textheight]{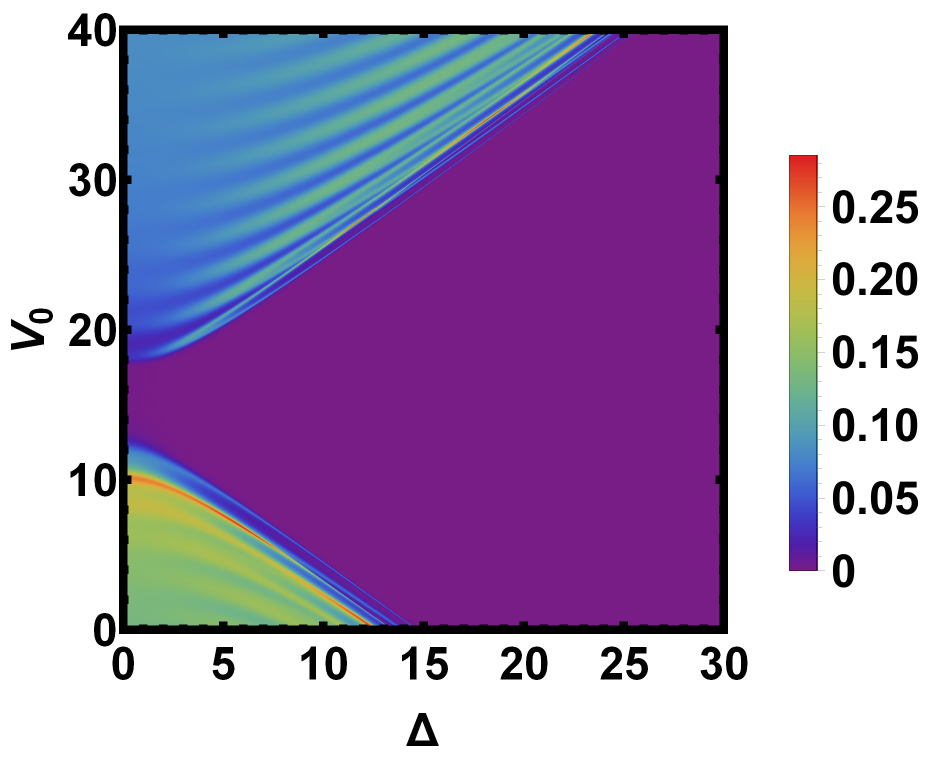}\label{fig10}}	\subfloat[]{\includegraphics[width=0.5\linewidth, height=0.165\textheight]{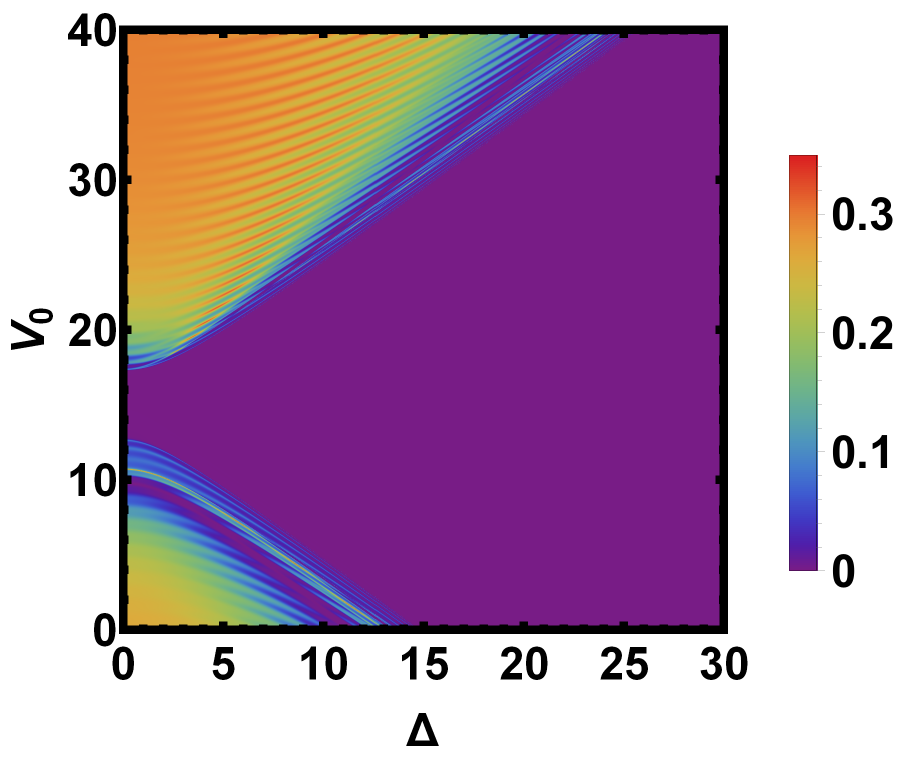}\label{fig12}}
		\caption{{Density plot of transmission probability for single photon emission $T_{-1}$ versus the  barrier height $V_0$ for $k_y=2$, $D=1.2$, $F=0.6$, $\omega=1$, $\varepsilon=15$. \text{{(a)}}\color{black}{:} Without strain $S=0.0$, \text{{(b)}}\color{black}{:} for zigzag direction $S=0,2$.}}\label{Np}
	\end{figure}
	
	To illustrate the effect of the applied strain on the transmission probability for single-photon emission, Fig.~\ref{Np} displays density plot of $T_{-1}$ versus the barrier height $V_0$ and the energy gap  $\Delta$ for $k_y=2$, $D=1.2$, $F=0.6$, $\omega=1$, $\varepsilon=15$. The results compare the unstrained case ($S=0.0$) with the zigzag-strained configuration ($S=0.2$). From Fig.~\ref{Np}(a), it is evident that the transmission increases significantly for low barrier heights $0\leq V_0\leq10$, while it drops sharply as the barrier becomes higher.  Moreover, $T_{-1}$ decreases rapidly as $\Delta$ increases, eventually becoming almost completely suppressed for the highest gap values  \cite{density}. This reduction occurs because increasing the gap blocks many propagating states and converts most modes in the barrier into evanescent ones. In contrast, when a zigzag strain of $S=0.2$ is applied  in Fig.~\ref{Np}(b), we observe that $T_{-1}$ increases significantly as $V_0$ grows, and the number of oscillations becomes larger. Additionally, one can clearly see the emergence of pronounced transmission peaks.  This happens because the zigzag strain changes the electronic structure and shifts the allowed propagating states inside the barrier. As a result, the correspondence between the incoming wave and the wave inside the barrier improves over a wider range of $V_0$ and increases the probability of transmission. The strain also changes the phase of electrons as they cross the barrier, leading to stronger interference effects. This creates more oscillations  and clear resonance peaks whenever the interference is constructive. Finally, it can thus be concluded that the deformation is more effective when both a gap and an applied potential are present.

	Fig.~\ref{r} presents how strain amplitude  along zigzag direction affects the transmission probability for single photon emission $T_{-1}$, for parameters  $k_y=2$, $D=1.2$, $F=0.6$, $\omega=1$,  $\varepsilon=15$. Fig.~\ref{r}(a) corresponds to $V_0=2$ (blue line), $V_0=4$ (green line), $V_0=6$ (red line) at $\Delta=3$, while Fig.~\ref{r}(b) corresponds to $\Delta=1$ (blue line), $\Delta=3$ (green line), $\Delta=5$ (red line) at $V_0=4$. Indeed, we examine the transmission across three distinct regions. For the region $0<S<0.11$,  $T_{-1}$ appears at different points and increases as $V_0$ rises,  showing the appearance of quantum interference effects  due to a slight shift in the propagating states.  Beyond this range, particularly in interval $0.11<S<0.22$,  $T_{-1}$ displays oscillatory behavior and gradually decreases as $V_0$ increases. At larger deformation, $T_{-1}$  shows sharp peaks under specific conditions but ultimately vanishes as shown in \cite{M1,D2}.
	\begin{figure}[H]\centering
		\subfloat[]{\includegraphics[width=0.5\linewidth, height=0.215\textheight]{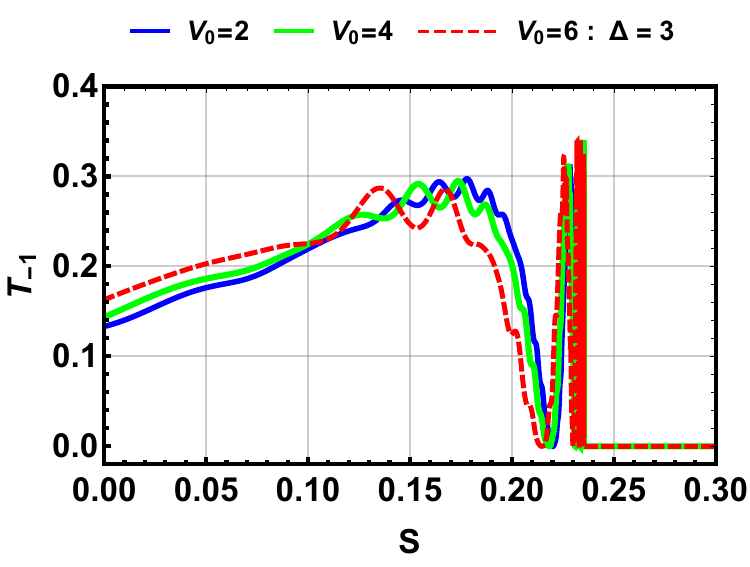}\label{yhy}}	\subfloat[]{\includegraphics[width=0.5\linewidth, height=0.215\textheight]{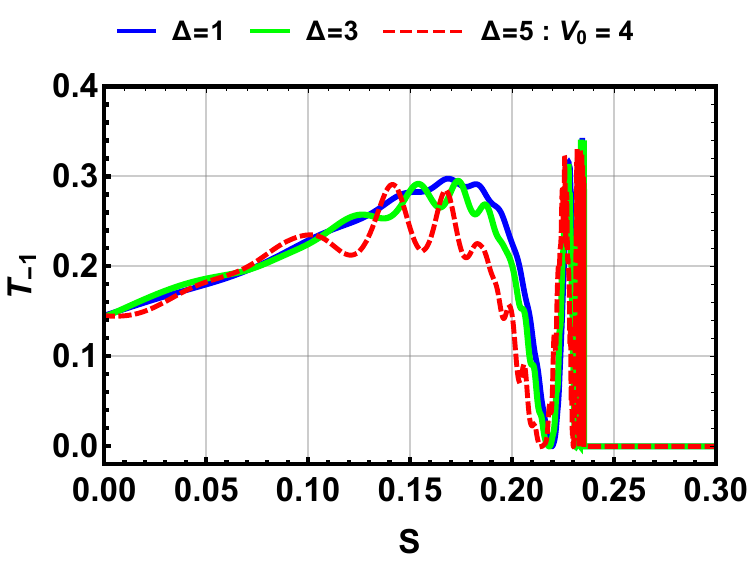}\label{yh}}	
		\caption{{Transmission probability for single photon emission $T_{-1}$ versus the zigzag strain $S$ for  $k_y=2$, $D=1.2$, $F=0.6$, $\omega=1$,  $\varepsilon=15$. \text{{(a)}}\color{black}{:} $V_0=2$ (blue line), $V_0=4$ (green line), $V_0=6$ (red line), $\Delta=3$. \text{{(b)}}\color{black}{:} $\Delta=1$ (blue line), $\Delta=3$ (green line), $\Delta=5$ (red line), $V_0=4$.}}\label{r}
	\end{figure}

	Fig.~\ref{Nmi} shows the transmission $T_{+1}$ versus barrier height $V_0$ for single-photon absorption, with $k_y=2$, $D=1.2$, $F=0.6$, $\omega=1$, and $\Delta=3$. In Fig.~\ref{Nmi}(a), the effect of zigzag strain is shown at fixed energy $\varepsilon=15$ for $S=0.0$ (blue line), $S=0.1$ (green line), and $S=0.2$ (red line).
	%
	%
	In Fig.~\ref{Nmi}(b), the effect of the incident energy is highlighted for a fixed strain $S=0.2$, with  $\varepsilon=10$ (blue line), $\varepsilon=15$  (green line), and $\varepsilon=20$ (red line). 
	The analysis of Fig.~\ref{Nmi}(a) shows that increasing zigzag strain enhances $T_{+1}$ and produces a larger number of oscillations \cite{chnafa1,M1,chnafa3}. This behavior results from strain-induced modifications of the barrier’s band structure, which improve wavefunction matching, enhance tunneling, and reinforce interference effects, leading to more pronounced transmission oscillations.
	In contrast, Fig.~\ref{Nmi}(b) shows that as the incident energy increases, $T_{+1}$ shifts to higher $V_0$ values and eventually stabilizes, remaining nearly constant ($T_{+1}= 0.3$) for the three chosen energies. Physically, at low energies, the band gap strongly restricts the available propagating states, reducing the transmission amplitude. However, as $\varepsilon$ increases, electrons acquire sufficient kinetic energy to cross the barrier, making the transmission largely independent of the gap.

	\begin{figure}[H]\centering
		\subfloat[]{\includegraphics[width=0.5\linewidth, height=0.22\textheight]{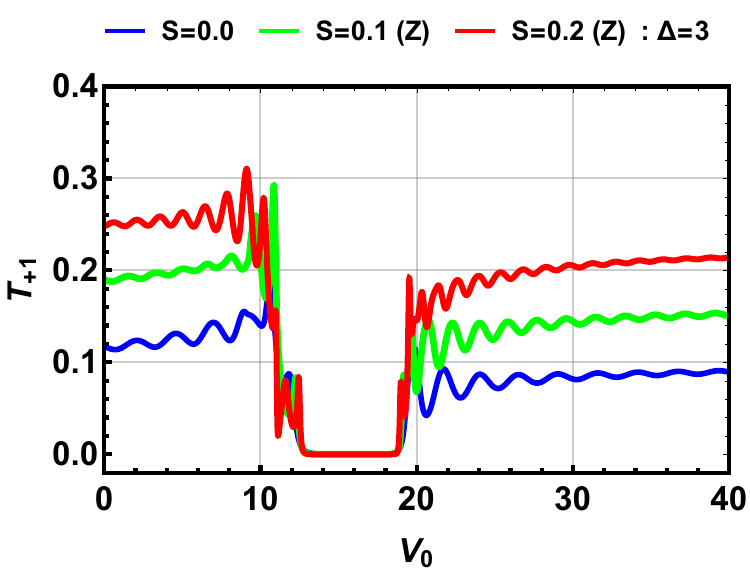}\label{fig13}}	\subfloat[]{\includegraphics[width=0.5\linewidth, height=0.22\textheight]{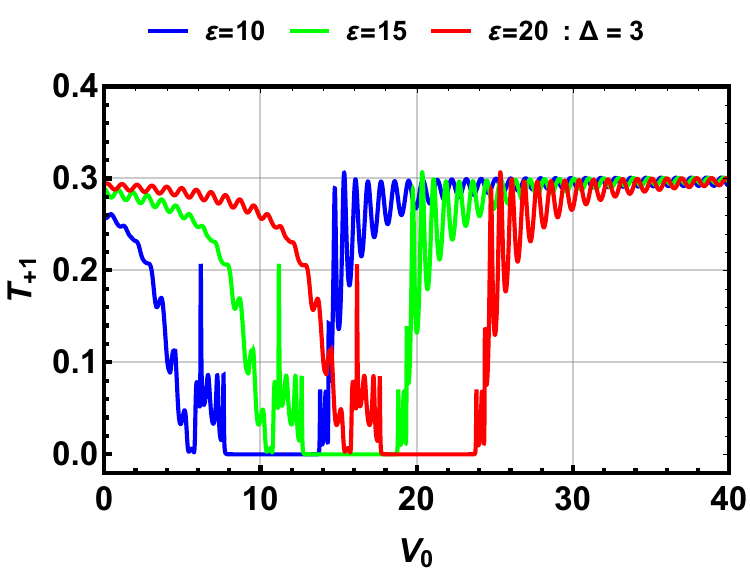}\label{fig15}}
		\caption{{Transmission probability for single photon absorption $T_{+1}$ versus the  barrier height $V_0$ for $k_y=2$, $D=1.2$, $F=0.6$, $\omega=1$, $\Delta=3$.  \text{{(a)}}\color{black}{:} $\varepsilon=15$,  $S=0.0$ (blue line), $S=0.1$ (Z) (green line), $S=0.2$ (Z) (red line).  \text{{(b)}}\color{black}{:} $S=0.2$,  $\varepsilon=10$, (blue line), $\varepsilon=15$  (green line), $\varepsilon=20$ (red line).}}\label{Nmi}
	\end{figure}
	Fig.~\ref{p} presents a density plot of the transmission probability for single photon absorption $T_{+1}$ versus the  barrier height $V_0$ and the barrier width $D$ with $k_y=2$, $F=0.6$, $\omega=1$, $\varepsilon=15$, $\Delta=3$. We observe that $T_{+1}$ exhibits an oscillatory behavior whose amplitude and distribution change as $D$ increases, showing similar behavior as in silicene \cite{3}. In Fig.~\ref{p}(a) for ($S=0.0$),  the oscillations appear more regular, with transmission maxima and minima following a relatively uniform pattern as both $V_0$ and $D$ vary. However, when a zigzag strain $S=0.2$ is applied, as illustrated in Fig.~\ref{p}(b), the transmission profile undergoes a substantial modification. The oscillations become more pronounced and their amplitudes increase noticeably.  This difference can be explained by the fact that the strain modifies the structure of the electronic bands inside the barrier, thereby improving the coupling between the electrons.  This leads to enhanced tunneling probabilities and stronger interference effects, which amplify the oscillatory features. Consequently, comparing the two cases clearly demonstrates that strain plays a crucial role in shaping the transmission profile.
	
	\begin{figure}[H]\centering
		\subfloat[]{\includegraphics[width=0.5\linewidth, height=0.17\textheight]{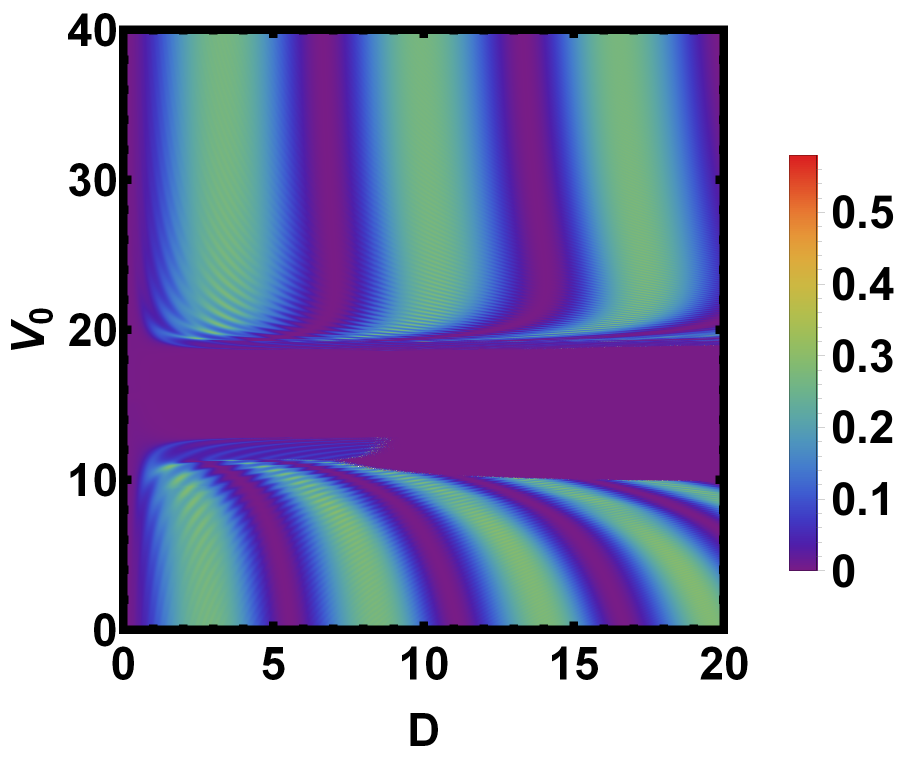}\label{fg10}}	\subfloat[]{\includegraphics[width=0.5\linewidth, height=0.17\textheight]{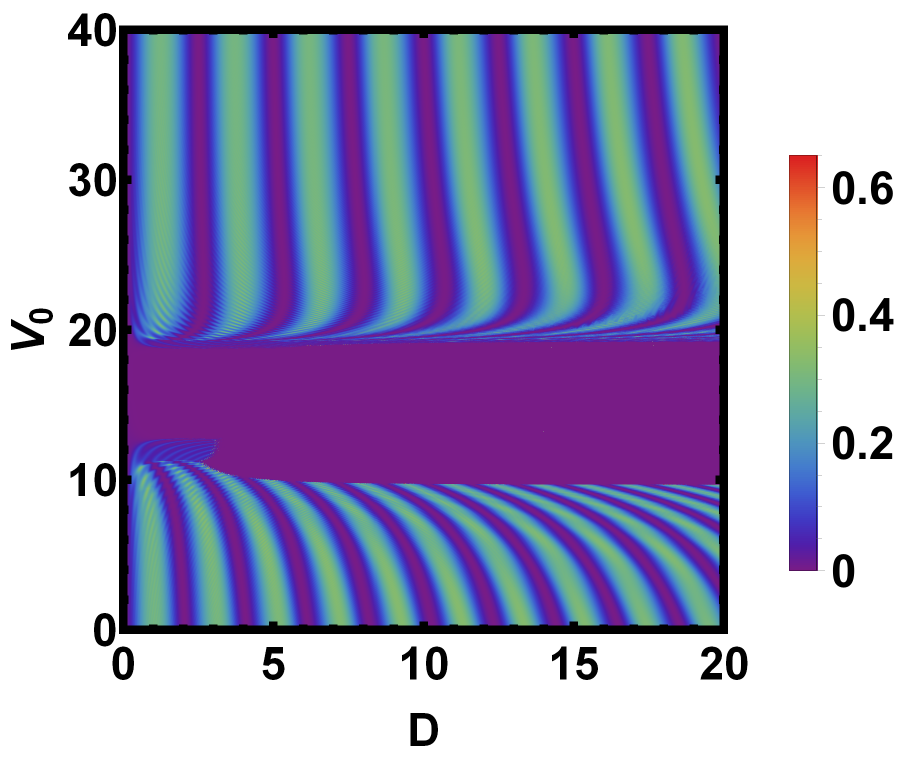}\label{fi}}
		\caption{{Density plot of transmission probability for single photon absorption $T_{+1}$ versus the  barrier height $V_0$ and the barrier width $D$ for $k_y=2$, $F=0.6$, $\omega=1$, $\varepsilon=15$, $\Delta=3$. \text{{(a)}}\color{black}{:} Without strain $S=0.0$, \text{{(b)}}\color{black}{:} for zigzag direction $S=0,2$.}}\label{p}
	\end{figure}

	\begin{figure}[H]\centering
		\subfloat[]{\includegraphics[width=0.5\linewidth, height=0.215\textheight]{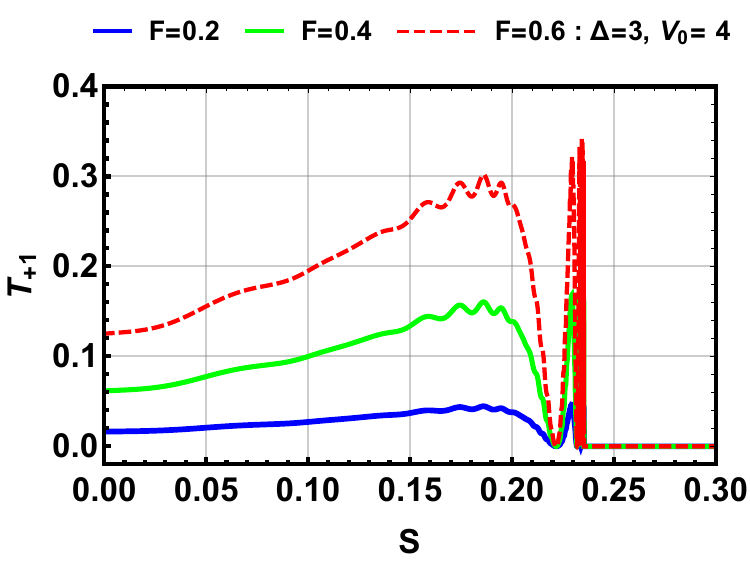}\label{fg0}}	\subfloat[]{\includegraphics[width=0.5\linewidth, height=0.215\textheight]{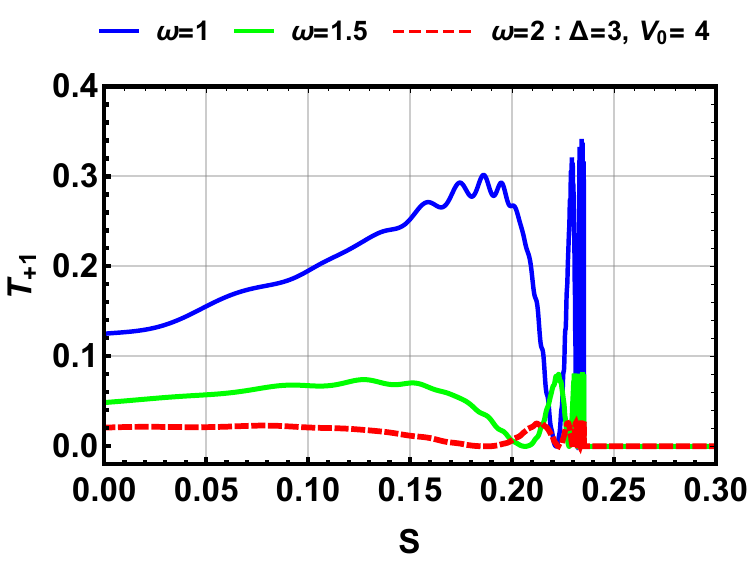}\label{fiu}}
		\caption{{Transmission probability for single photon absorption $T_{+1}$ versus the zigzag strain $S$ for  $k_y=2$, $D=1.2$,  $\varepsilon=15$, $\Delta=3$, $V_0=4$. \text{{(a)}}\color{black}{:} $F=0.2$ (blue line), $F=0.4$ (green line), $F=0.6$ (red line), $\omega=1$. \text{{(b)}}\color{black}{:} $\omega=1$ (blue line), $\omega=1.5$ (green line), $\omega=2$ (red line), $F=0.6$.}}\label{ha}
	\end{figure}
	
	Fig.~\ref{ha} illustrates the effect of the laser field amplitude $F$ and frequency $\omega$ on the single-photon absorption transmission $T_{+1}$ versus the zigzag strain $S$, with parameters $k_y=2$, $D=1.2$, $\varepsilon=15$, $\Delta=3$, and $V_0=4$. In Fig.~\ref{ha}(a), we consider $\omega=1$ and $F=0.2, 0.4, 0.6$, observing a monotonic increase of $T_{+1}$ with $F$, consistent with \cite{chnafa1,chnafa2,El Aitouni}. This indicates that applying a zigzag deformation enhances the coupling between Dirac electrons and the oscillating laser field, increasing the probability of photon absorption and tunneling through the barrier.
	In Fig.~\ref{ha}(b), we fix $F=0.6$ and vary the frequency as $\omega=1, 1.5, 2$. We observe that $T_{+1}$ decreases with increasing $\omega$, in agreement with \cite{chnafa1,chnafa2,El Aitouni,M1,D2}. This behavior occurs because higher frequencies reduce the interaction time between electrons and the oscillating field, lowering the probability of single-photon absorption and tunneling through the barrier. These results demonstrate that photon-assisted transport in strained graphene can be effectively controlled by tuning both mechanical strain and electromagnetic field parameters.
	

	\section{Conclusion}\label{TSFOR2}
	
	The transport of massive Dirac particles through strained barriers under a linearly polarized laser field and scalar potentials has been extensively studied. These studies assume that the strain is applied uniaxially along the zigzag direction of the graphene sheet. The Floquet approximation was employed to account for the temporal periodicity of the laser field. Continuity of wavefunctions across three regions yielded equations containing an infinite number of modes corresponding to the laser-induced subbands. To handle this complexity, we employed a matrix formalism that enabled us to compute all transmission amplitudes. Using the current densities, we calculated the transmission probabilities for each energy band. 
	To simplify the analysis, we focused 
	on three transmission modes: the central band ($m=0$) and the first sidebands corresponding to photon absorption ($m=+1$) and emission ($m=-1$).
	

	Next, we conducted a numerical analysis of the transmission probabilities to examine the influence of zigzag strain, energy gap, scalar potential, and laser parameters. We found that in the absence of strain, transmission exhibits oscillatory behavior in all modes. By varying the energy gap and scalar potential, we observed that transmission through the central band and the first photon-emission sideband decreased with varying amplitudes. Conversely, when strain is applied along the zigzag direction, the Fano-type oscillations become more pronounced but gradually disappear as strain values increase. Another notable finding is that the transmission profile increases at $\varepsilon<V_0$ but decreases at $\varepsilon>V_0$. Furthermore, we demonstrated that transmission via photon absorption is significantly impacted by the incidence energy, shifting the resonance peaks to the right and becoming stable at higher $V_0$ values.   Additionally, increasing the barrier width produces distinctive oscillatory patterns with sharp resonant peaks. We revealed that transmission slightly increased as $F$ varied, and increased or decreased drastically as $S$ amplified.
	
	
	Our study shows that electron transport in graphene can be controlled by combining uniaxial zigzag strain, an energy gap, a scalar potential, and laser irradiation. Adjusting these parameters precisely tunes the transmission properties to achieve the desired electronic behaviors. Based on experiments \cite{N,E}, our outcomes can be reproduced. These experiments indicate that this combination provides a flexible platform for controlling electron transport in graphene, opening possibilities for advanced electronic and optoelectronic devices.
	

\end{document}